%% file: 2018-issta-thaller-gradient-docSym.tex
\documentclass[a4paper,USenglish]{lipics}

\usepackage{microtype}
\usepackage{setspace}
\usepackage{amsmath,amssymb,amsfonts, bm, amsthm}
\usepackage{algorithmic}
\usepackage{graphicx, tikz}
\usepackage{booktabs, multirow}
\usepackage{stfloats}
\usepackage[inline]{enumitem}
\usepackage{url}
\usepackage{xcolor}
\usepackage{caption}
\usepackage{hyperref}
\usepackage{pifont}
\usepackage[framemethod=tikz]{mdframed}

\graphicspath{{images/}}
\DeclareGraphicsExtensions{.pdf,.jpeg,.png}

\definecolor{dark-red}{rgb}{.6, .15, .15}
\definecolor{dark-blue}{rgb}{.15, .15, .55}
\definecolor{accent1}{HTML}{24B9FC}
\definecolor{accent2}{HTML}{9ECD67}
\definecolor{accent3}{HTML}{FFA929}
\definecolor{light-bg}{HTML}{B2B2B2}

\hypersetup{
	colorlinks,
  linkcolor={dark-red},
	citecolor={dark-blue},
  urlcolor={black}
}

\newcommand{\code}[1]{\texttt{#1}}
\newcommand{\circleOne}{\ding{192}}
\newcommand{\circleTwo}{\ding{193}}
\newcommand{\circleThree}{\ding{194}}
\newcommand{\circleFour}{\ding{195}}
\newcommand{\circleFive}{\ding{196}}
\newcommand{\circleSix}{\ding{197}}
\newcommand{\circleSeven}{\ding{198}}

\graphicspath{{images/}}

\title{Probabilistic Software Modeling\footnote{The research reported in this paper has been supported by the Austrian Ministry for Transport, Innovation and Technology, the Federal Ministry of Science, Research and Economy, and the Province of Upper Austria in the frame of the COMET center SCCH.}}

\author[1]{Hannes Thaller}
\affil[1]{Institute for Software Systems Engineering \\Johannes Kepler University Linz, Austria\\
  \texttt{hannes.thaller@jku.at}\\
  ~\\
  Advised by Alexander Egyed (\texttt{alexander.egyed@jku.at}) \\
  Co-Advised by Lukas Linsbauer (\texttt{lukas.linsbauer@jku.at})
}
\authorrunning{H. Thaller} 

\Copyright{Hannes Thaller}


\serieslogo{}
\volumeinfo
  {Billy Editor and Bill Editors}
  {2}
  {Conference title on which this volume is based on}
  {1}
  {1}
  {1}
\EventShortName{}
\DOI{10.4230/LIPIcs.xxx.yyy.p}

\begin{document}

\maketitle

\begin{abstract}
Software Engineering and the implementation of software has become a challenging task as many tools, frameworks and languages must be orchestrated into one functioning piece.
This complexity increases the need for testing and analysis methodologies that aid the developers and engineers as the software grows and evolves.
The amount of resources that companies budget for testing and analysis is limited, highlighting the importance of automation for economic software development.
We propose Probabilistic Software Modeling, a new paradigm for software modeling that builds on the fact that software is an easy-to-monitor environment from which statistical models can be built.
Probabilistic Software Modeling provides increased comprehension for engineers without changing the level of abstraction.
The approach relies on the recursive decomposition principle of object-oriented programming to build hierarchies of probabilistic models that are fitted via observations collected at runtime of a software system.
This leads to a network of models that mirror the static structure of the software system while modeling its dynamic runtime behavior.
The resulting models can be used in applications such as test-case generation, anomaly and outlier detection, probabilistic program simulation, or state predictions.
Ideally, probabilistic software modeling allows the use of the entire spectrum of statistical modeling and inference for software, enabling in-depth analysis and generative procedures for software.
 \end{abstract}

\input{main}

\setstretch{1}
\bibliography{psm}

\end{document}

%% file: main.tex
\section{Introduction}

Software and software development is an omnipresent and growing industry, and many modern advances in other domains hinge on the success in computer science.
Many software systems never stop growing and evolving, consuming thousands of developer days; hence, knowledge and understanding of these systems are essential.
This growth poses new requirements in software engineering, not only on an organizational level but also on a functional level.
How can we understand the behavior of a system?
How can we apply reuse on a broad scale?
How can we guarantee a specific behavior?
How can we automate development efforts?
Many of these questions are already (partially) answered by existing modeling techniques while others are not.

The two main modeling approaches currently are Model-Driven Engineering (MDE) \cite{Schmidt2006} and Formal Methods (FM) \cite{Clarke1996}, which share some commonalities but have different goals.
In MDE, engineers create domain models on a higher abstraction level than source code to scale the overall development process.
A typical technique is to create a Domain Specific (Visual) Language (DSL) from which code and reference implementations can be generated and reused.
Another prominent MDE example is the Unified Modeling Language (UML) \cite{Rumbaugh2004} that conceptualizes many aspects of software development.
Formal Methods focuses on software specifications via logic.
Resulting models are unambiguous machine interpretable specifications used to generate the application code.
The resulting code needs less testing since many program states and transitions can be verified by automatic proofs.
Relative to MDE, FM lowers the abstraction level as concepts such as conditions and invariants, need to be designed, elicited and foremost proven before a single line of code is executed.
Many other specialized modeling techniques stem from these two approaches, e.g., model-based testing borrows techniques from formal methods.
While these modeling techniques offer significant benefits in their respective areas, both lack a way to inspect software and its runtime behavior in a direct, visual and interactive way an engineer would appreciate.
MDE lifts the abstraction level on which engineers reason about the best structure for reuse, while FM lowers the abstraction level down into the regime of logic where no ambiguity is allowed.

\begin{figure}[ht]
	\centering
	\includegraphics[width=.45\textwidth]{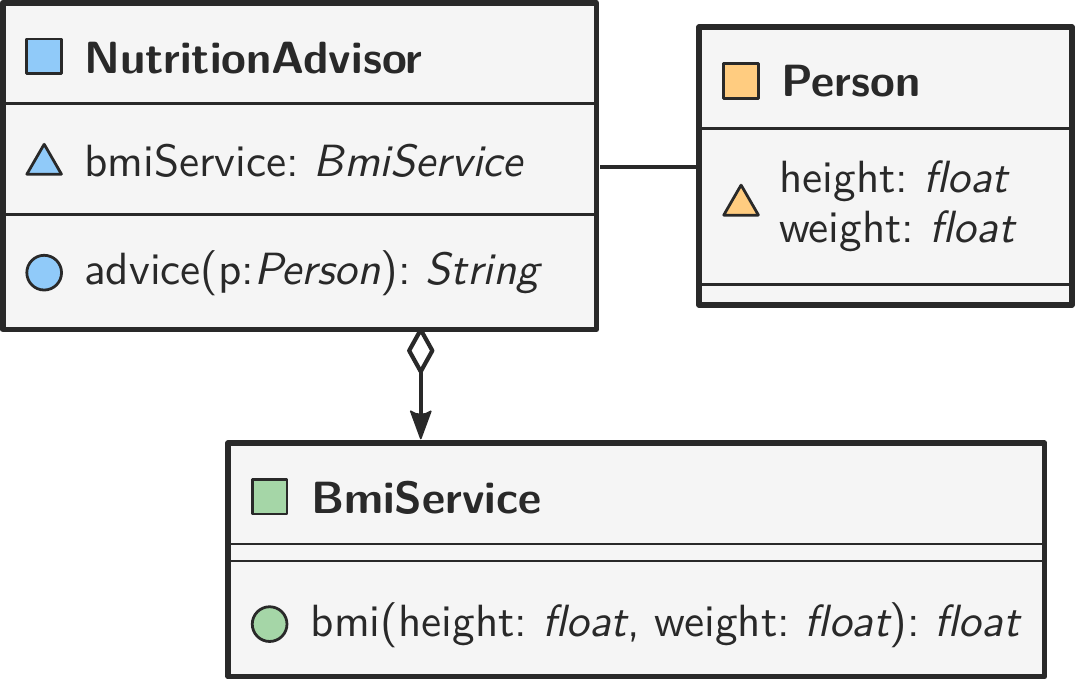}
	\caption{The class diagram is depicting the structure of a Nutrition Advisor system.
	\code{NutritionAdvisor} is the main class and entry point for requests.
	 Its \code{advice} method takes a \code{Person} entity and returns a textual advice regarding its nutrition.
	 \code{Person} stores nutrition relevant properties that \code{BmiService} service uses to compute the Body Mass Index (BMI) on which the textual advice is based.
	 }
	\label{fig:nutrition advisor structural}
\end{figure}

\begin{figure}[ht]
	\centering
	\includegraphics[width=.45\columnwidth]{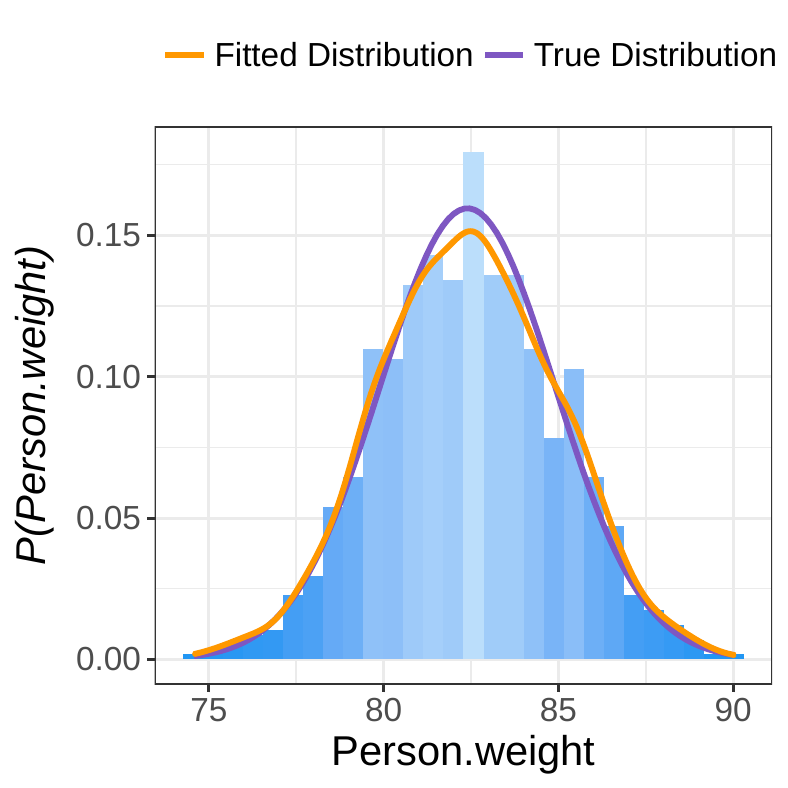}
	\caption{Distribution of the $Weight$ variable of the $Person$ class depicting its behavior at runtime.
	The histogram shows all \code{weight} values that objects of type \code{Person} had during the runtime.
	These values were generated using the \emph{True Distribution} (purple) and approximated by the \emph{Fitted Distribution} (yellow).
	}
	\label{fig:nutrition advisor prob behavioral}
\end{figure}

We propose \emph{Probabilistic Software Modeling (PSM)}, a data-driven modeling paradigm for software that allows simulation and analysis of software systems.
PSM builds a \emph{Probabilistic Model (PM)} from the system under inspection by analyzing its structure, i.e., types, properties and executables, and behavior via static and dynamic code analysis.
These probabilistic models are then used in typical predictive and generative tasks known from statistical inference.
For example, the visualization of a property's distribution at runtime (e.g., \code{Person.weight} in Figure \ref{fig:nutrition advisor structural} with behavior in Figure \ref{fig:nutrition advisor prob behavioral}), or prediction of the next most likely value (or distribution of values) of a property's active state, etc.
Furthermore, inference allows the simulation of entire sub-systems through probabilistic execution of the corresponding models.

PSM positions itself as a complementary software modeling approach to MDE or FM.
It offers engineers and programmers to interact with the software's structure and behavior through statistical models.
Forward and Lethbridge \cite{Forward2008} showed that many issues with code-centric comprehension and analysis, i.e., fixing bugs, analyzing structure and behavior by inspecting the source code, come from the fact that developers cannot see and feel the structure and behavior of their system, especially after changes.
This results in a typical behavior where programmers, even if they are confident in their code through tests, tend to inspect changes in code via debugging to see an instantiation of its runtime behavior.
PSM tries to alleviate these issues by making structure and behavior explicit in models.
The usage patterns of PSM are focused on increasing confidence through comprehension and white-boxing of software.
PSM provides fine-grained control of variables and their interactions within a system, letting the engineers see and manipulate what happens at runtime through PMs and statistical inference.
Idealistically, PSM allows companies that provide software components such as APIs, plugins, frameworks, to exchange their probabilistic models with partners.
These models are then used to inspect and check the behavior, generate tests from the models, or to simulate and infer the behavior at the boundaries of internal and external components via the PMs of the partners.
That is a unified communication framework via models without the need for a full running system.
Again, PSM is a complementary approach to software modeling, and existing practices should benefit from it.
MDE can be paired with PSM to check the compatibility and integrity of a new configuration build via MDE.
FM could benefit from PSM by using it to check and generate tests from non-critical sub-systems in a less rigorous and data-driven way.
In sum, PSM is a modeling paradigm for the analysis of systems on the abstraction level of code, hence, complements existing techniques in their effort for better software comprehension.

\section{Applications}\label{sec:applications}
Probabilistic software modeling is, to the best of our knowledge, the first attempt to model the software's structure and behavior purely in terms of probabilistic models.
Hence, it might not be clear what applications PSM can enable.
As mentioned, PMs allow for \emph{predictive} and \emph{generative} tasks, however, this discrimination is purely conceptual in the context of PSM as the underlying techniques (e.g., approximate inference by sampling \cite{Murphy2012}) and their associated notations are the same.

Predictive tasks use the PMs to predict probabilities, values, and distributions according to criteria in an inferential query.
For instance, given the \code{Person} class and its properties in Figure \ref{fig:nutrition advisor structural}, it is possible to query the PM how likely a specific value or value range is, e.g.,  $P(Weight > 80.0)$.
By conditioning on other variables, it is possible to infer relationships between variables.
For example, $P(Weight \mid 169.0 < Height < 170.0)$ results in the distribution of $Weight$ values for persons with a height of $169cm$.

Generative tasks sample observations from the PMs underlying distributions by either using the distribution as-is or by conditioning on other variables.
For instance, the PM can be used to instantiate objects of the \code{Person} class by sampling for its properties $P(Height, Weight) \sim (69.54, 169.59)^T$.
Similar to the predictive tasks, it is possible to sample values from conditioned distributions, effectively generating samples according to some criteria.
Both cases are inferential tasks that allow a broad range of possible applications.
The following non-exhaustive list are applications that are part of the future research.

\subsection{Predictive Applications}
Predictive applications seek to quantify, visualize and analyze the behavior of software.

\textbf{Visualization \& Comprehension} applications of PSM have the goal to white-box software by making the runtime explicit for engineers.
The visualizations show the distribution of properties, inputs, and outputs of executables.
These visualizations can be adapted by conditioning the underlying source models with specific variables.
For example, what does the Body Mass Index (BMI) look like for people with a height of $160cm$?
Furthermore, untypical examples can be visualized and followed through other models effectively visualizing how far they ripple through the system and the effect they have on downstream components.
For instance, how far can an invalid weight of $-10$ ripple through the application without being caught and which effect does it have?

\textbf{Integrity \& Compatibility Evaluation} quantifies the consistency between and within software components.
Integrity measures the amount of divergence between models of different versions of a component.
The divergence is the integrity measure aggregating the distribution of accepted and unaccepted values. This distribution can be visualized, conditioned and used for further inferential tasks, e.g., spotting regression and analyzing expected behavior changes.
Compatibility evaluation measures the amount of divergence between models of different components and enables the same applications as integrity evaluation such as quantification, visualization, inferential queries.

\textbf{Anomaly Detection} \cite{Chandola2009, Hangal2002} applications can be employed during development but also in live systems.
A PM is used as reference behavior to check whether the newly observed behavior lies within the bounds of its distributions.
The implementation might be as simple as the inferential query $P(Weight=x) < 0.1$ where $x$ is the currently observed value.
The condition triggers an action if the likelihood of the current value is smaller then $0.1$.
Additionally, PSM can compute the ripple distance\footnote{Distance from the anomaly origin to the first time it is actively perceived, counted in functions calls (call stack).} and effect of the anomaly for detailed reports that summarize the incident.

\subsection{Generative Applications}
Generative applications reproduce the system's runtime behavior by generating new instances of code elements such as classes, parameters, and return values.

\textbf{Test-case generation} \cite{Ali2010, Cavarra2002} can be realized by sampling from the PMs.
For example, test-suites can be organized into typical, rare, and impossible where the categories indicate the probability range the sampled values are allowed to have.
A test-suite with rare invocations for the \code{NutritionAdvisor.advice(p:Person)} could be constructed by sampling values that with $ 0.02 < P(Person) = P(Height, Weight) < 0.1$ from the types PM.
This allows to generate tests regarding the prevalence of states and values providing a better understanding of the behavior.
Furthermore, test-case generation can be used to evaluate the integrity and compatibility of components partially.
For integrity evaluation, the PM of the old version generates tests for the new version to find regressions.
For compatibility evaluation, the call-site PM generates tests for the receiving site to find incompatibilities.

\textbf{Simulation}, similar to test-case generation, samples values from the PMs and conditions downstream PMs with the sampled values.
This effectively executes the system in a probabilistic fashion with the expected behavior according to the underlying distributions.

\section{Approach}
Probabilistic Software Modeling builds on the fact that software systems and programming languages are based on the recursive decomposition principle \cite{Palmer1986}.
\emph{Any complex information event on one level can be described on another level by a number of components and a process specifying the relations among these.}
Reinterpreted for PSM and modern object-oriented software development: Instead of modeling the entire software (structure and behavior) in one complex probabilistic model, a recursive decomposition it into smaller models and their relationships yields a model with similar behavior.
For that, the level of abstraction needs to be fixed.
As mentioned, PSM is designed to analyze software on the same abstraction level developers are used to making, types, properties, and executables an appropriate choice.

\begin{figure*}[ht]
	\centering
	\includegraphics[width=.9\textwidth]{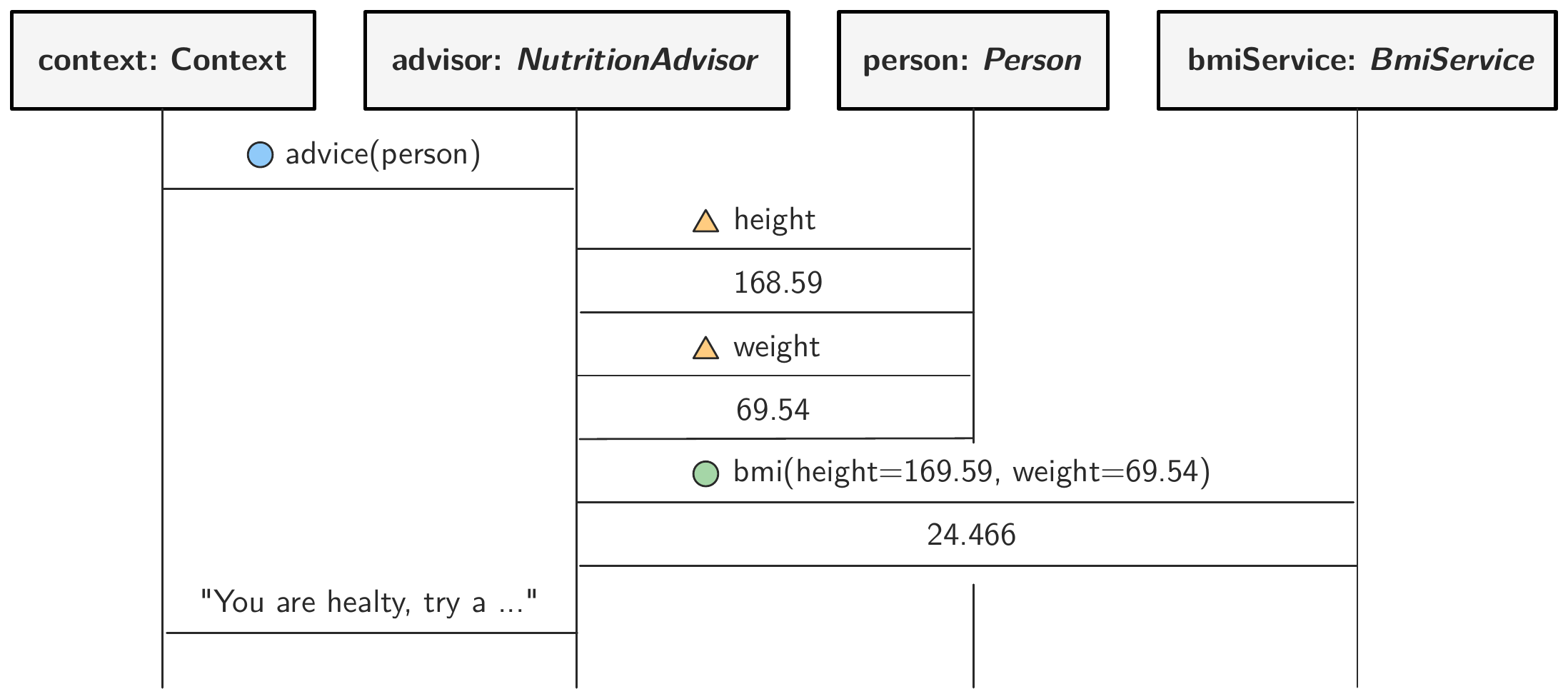}
	\caption{Example request to the Nutrition Advisor depicting its behavior.
		The \code{client} makes a new request which is handled via the \code{advisor} method of the \code{NutritionAdvisor} class.
		The method retrieves the properties of the \code{person} object and computes the BMI using the \code{bmiService}.
		Based on the result it returns a textual advice.
	}
	\label{fig:nutrition advisor behavioral}
\end{figure*}

Figure \ref{fig:nutrition advisor structural} shows an example of a nutrition advisor application.
\code{NutritionAdvisor} takes a \code{Person}, extracts its \code{height} and \code{weight} and feeds it into the \code{BmiService.bmi(.)} method that computes the BMI of a person.
Based on the result, it returns a textual advice regarding its nutrition.
Figure \ref{fig:nutrition advisor behavioral} illustrates one request to this application in the form of a sequence diagram.
In this example, the person is $168.59cm$ tall and $69.54kg$ heavy resulting in a BMI of $24.466$ by which the textual advice is computed.
A practical and straightforward application which exists (at least up to the advice) in various forms on the Internet.

\subsection{Usage Scenario}

\begin{figure*}[ht]
	\centering
 	\includegraphics[width=.93\textwidth]{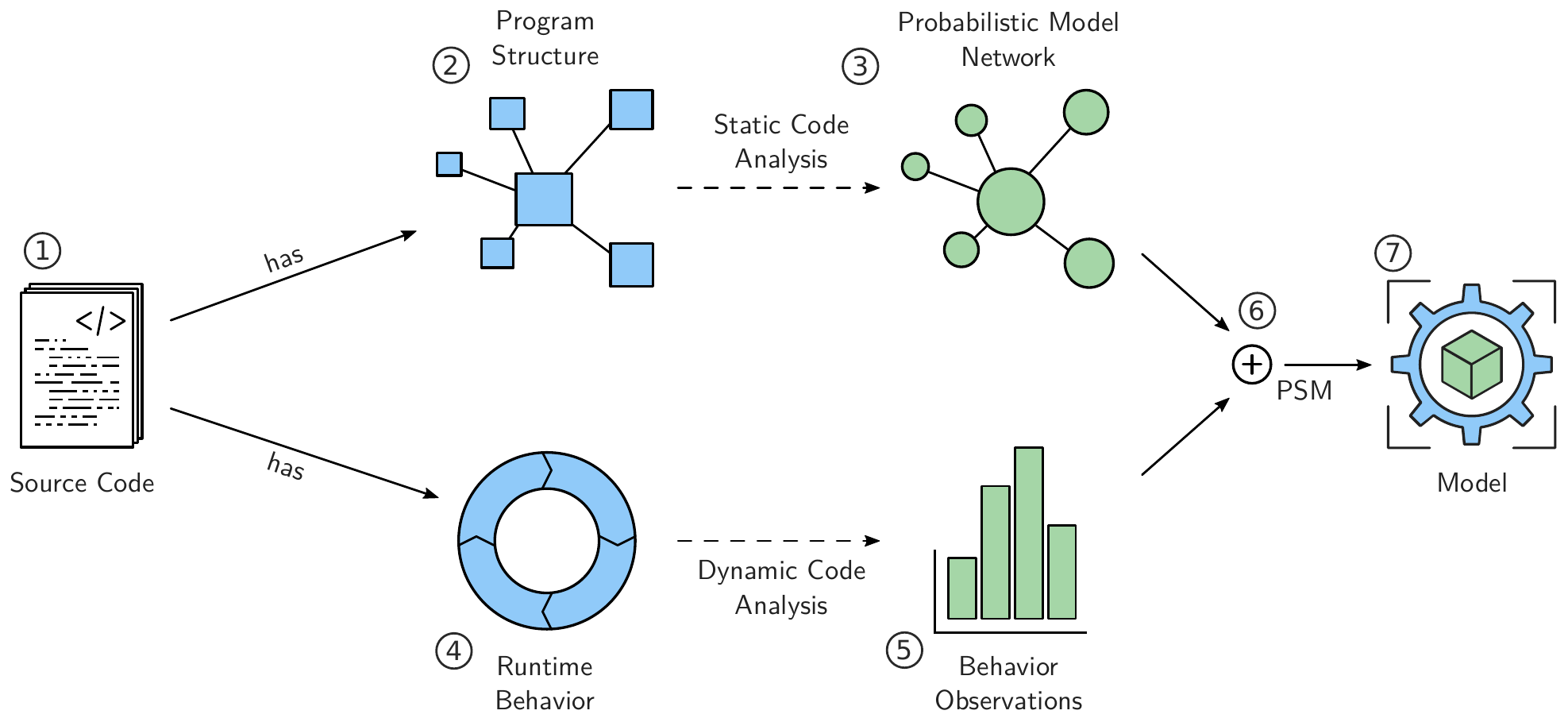}
	\caption{Source Code \circleOne{} has some structural \circleTwo{} and behavioral \circleFour{} properties that are extracted via static and dynamic code analysis.
	This results a network of probabilistic models \circleThree{} and behavior observations \circleFive{} that are combined by PSM \circleSix{} into the final model \circleSeven{}.
	}
	\label{fig:approach overview}
\end{figure*}

PSM is essentially a two-step approach in which we
\begin{enumerate*}[label=(\arabic*)]
	\item analyze the structure of the application via static code analysis, and then
	\item the behavior via dynamic code analysis (i.e., runtime monitoring).
\end{enumerate*}
Figure \ref{fig:approach overview} illustrates this basic usage scenario.
Starting with the \emph{Source Code} \circleOne{} that has some \emph{Structure} \circleTwo{}, we first analyze these static structural aspects (i.e., properties, types, executables) to define a \emph{Probabilistic Model Network} \circleThree{} that mimics this structure.
After that, the source code is executed to trigger a \emph{Runtime Behavior} \circleFour{} which provides the \emph{Behavior Observations} \circleFive{} via dynamic code analysis.
At last, PSM \circleSix{} approximates the behavior of each node within the probabilistic model network using the behavior observations to build the final \emph{Model} \circleSeven{}.
Figure \ref{fig:nutrition advisor prob behavioral} shows an example of such an approximated behavior by visualizing the distribution of \code{Person.weight} property in the Nutrition Advisor example.
The blue histogram is the monitored data from the dynamic code analysis while the yellow line is the fitted distribution.
In this running example, the data was sampled from the \emph{True Distribution} (purple), which is the target quantity (usually unknown) the PM approximates.
At this point, often the question arises whether it is possible to acquire a meaningful dataset.
Software is developed around data, and engineers have an excellent understanding of it.
That is, engineers solve problems, problems are represented by their data.
Hence, engineers need to have an understanding of the data to build a solution (problem-solving process \cite{Jewell1986, Sharp1991}).
A shop handles articles, a video-on-demand service offers videos, and the control system of a spaceship handles the remaining $\Delta v$, thus acquiring a meaningful dataset is often not a problem.

\begin{figure*}[ht]
	\centering
 	\includegraphics[width=.90\textwidth]{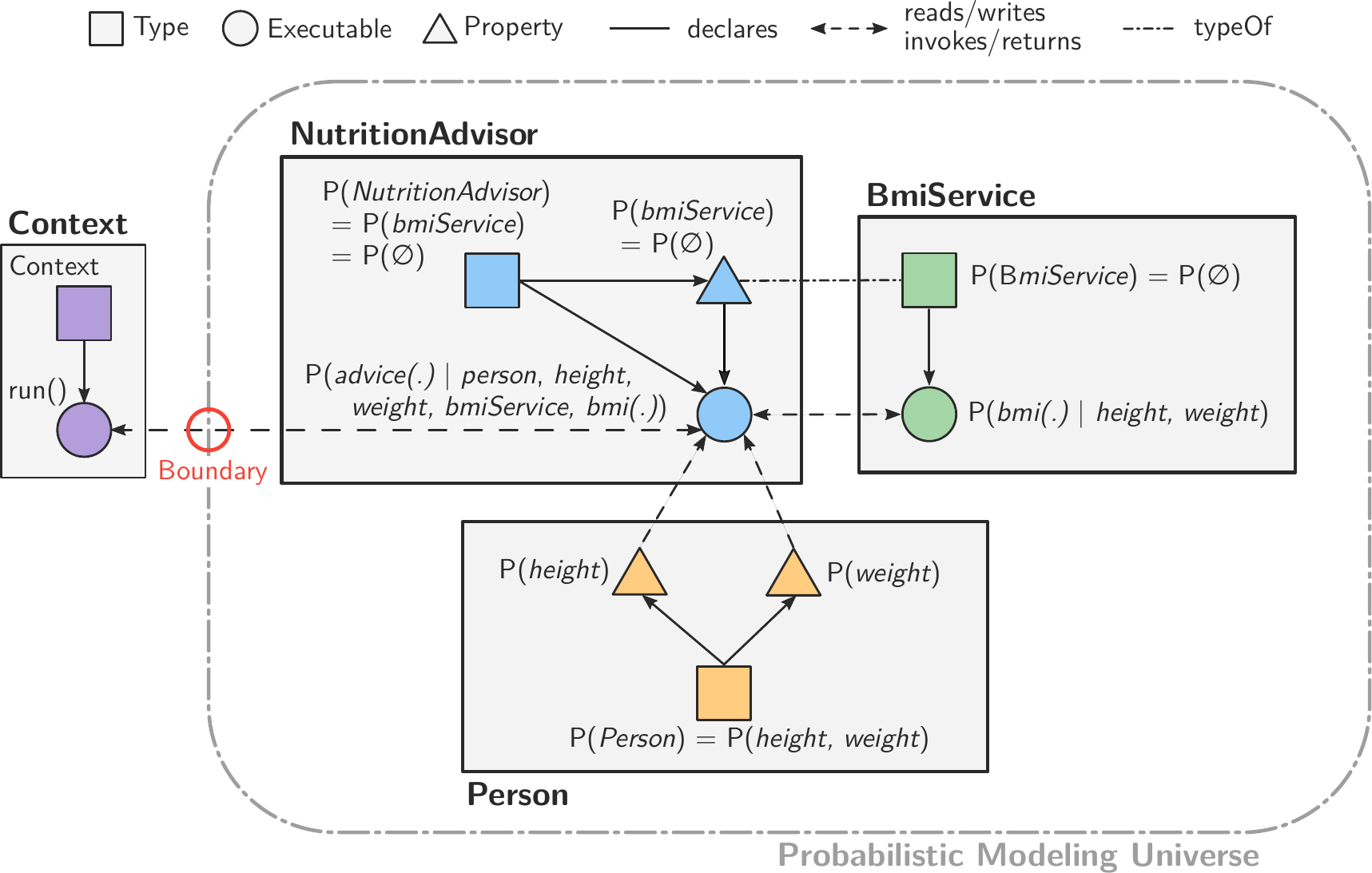}
	\caption{Graphical model of the Nutrition Advisor's structure of probabilistic models and random variables.
	A \emph{Property} is a model with one random variable representing the property itself.
	The joint distribution of all properties of a class represents the model of the enclosing \emph{Type}.
	An \emph{Executable} is a conditional model of its parameters, property reads, and method invocations.
	The \emph{Probabilistic Modeling Universe} are all classes that are considered during modeling.
	}
	\label{fig:nutrition advisor prob structural}
\end{figure*}

\subsection{Modeling}
Figure \ref{fig:nutrition advisor prob structural} shows the network of probabilistic models of the Nutrition Advisor example containing the random variables for types, properties, and executables.
Properties are random variables.
Types are quantified by their properties, hence, are joint models of their property's random variables.
In the case where a type does not have a property, the distribution of the type itself is empty.
The objects of a type that has no state are indistinguishable from each other; therefore there is no information to extract.
This applies transitively as the $bmiService$ property in the $NutritionAdvisor$ class shows.
Executables are random variables that are conditioned on their input parameters, property reads and method invocations (return values).
For instance, $advice(.)$ is conditioned on $person$ (parameter), $height, weight$ (property reads) and $bmi(.)$ (method invocation).
All classes that are considered by PSM are within the probabilistic modeling universe and can be used for the above-mentioned applications in Section \ref{sec:applications}.
However, there will always be external types beyond the modeling universe illustrated by the \code{Context} class in Figure \ref{fig:nutrition advisor prob structural}.
Boundaries between external and modeled code elements are bridged by interpreting the external elements as latent variables.
The effect of these variables on the PMs in the modeling universe is subject to the data and its distribution.
That is, no specific variables are constructed for external elements, but their dependencies and effects to universe-internal elements are described through the runtime data itself (statistical dependencies).

To summarize, PSM uses random variables to construct small PMs of types, properties, and executables that are connected by conditioning on related elements.
Each PM is a generative model capable of producing the distribution of the code element according to the dynamically observed values.

\subsection{Inference Method Requirements}
The PMs in Figure \ref{fig:nutrition advisor prob structural} can be implemented with various statistical modeling techniques.
However, the application context imposes specific requirements onto the modeling method.
Modeling techniques have to be \begin{enumerate*}
	\item decidable,
	\item computationally efficient,
	\item non-parametric,
	\item universal.
\end{enumerate*}
Decidable such that a clear exit criterion for the density approximation exists.
Computationally efficient such that large amounts of data and models can be fitted with a limited budget.
Non-parametric such that no assumptions about the data itself have to be made.
Universal such that any given distribution can be approximated.
The prototype will be based on Variational Autoencoders (VAEs) \cite{Kingma2013}, and Probabilistic Programming \cite{Gordon2014} as these were deemed most promising.

\subsection{Limitations}
PSM has several limitations and constraints.
One constraint is that subsystems need to be runnable.
There is no limitation on the size of a subsystem as models can be built from functions which are the smallest runnable piece of code.
However, the fact that it needs a runtime in itself may be a limitation.
Furthermore, the need for a sensible runtime extends this limitation.
PSM models reflect a program and enable probabilistic inference on them.
Hence, the fitted behavior will be similar to the program's behavior at runtime.
Running a program with arbitrary data reduces the PSM analysis capabilities to interactions of elements since the distributions of the elements themselves might not be representable for the real world.
At last, the programming language must be reasonably structured as models are on the level of executables, i.e., the recursive decomposition principle must apply to the programming language.
Whether this decomposition is enforced via classes and their objects, or via procedures is irrelevant.
Furthermore, the language must allow for runtime monitoring.

\section{Related Work}
The Unified Modeling Language (UML) \cite{Rumbaugh2004} is a prominent example of visual modeling of software.
Its primary purpose is to aid engineers in their design efforts by defining diagrams that capture module, sequence, requirement or state concerns.
Another goal of UML is communication by providing a unified visual representation of software design.

UML is often associated with Model-driven Engineering (MDE) \cite{Schmidt2006} as it can be used as a modeling language.
MDE is a modeling approach with a wide variety of techniques targeted for reuse and synthesis of software using UML or DSLs.
It achieves this by increasing the level of abstraction resulting in increased productivity, portability, maintainability and interoperability.
However, increasing the abstraction by introducing a new artifact adds mapping costs between the high-level abstraction and the low-level representations.
Often the (low-level) code and the (high-level) model gets desynchronized \cite{Forward2008}, thus reducing its usefulness.
Also, technical issues and bug fixes cannot be solved in models.
Compared to PSM, MDE helps to improve engineering tasks before the modeled code is written.
On the other hand, PSM aids engineers in understanding the actual instantiation of the system by letting them interact with its behavior.
The study from Forward and Lethbridge \cite{Forward2008} suggests synergetic effects between MDE and PSM in their endeavor of software comprehension.
MDE supports engineers in solving model-centric issues, i.e., design, module decomposition and reuse, while PSM offers support for code-centric issues such as behavior comprehension, regression analysis, and fixing of bugs, that are often tedious in the source code.

Formal Methods is a modeling approach that focuses on software specification in logical terms.
Once a system's behavior is fully specified by formal models, the entire application can be synthesized from it.
This, ideally, makes testing obsolete as the specification itself can be part of proofs.
However, the transition from the uncertain real world into the perfect mathematical world has several issues.
It is time-consuming, needs personnel with the appropriate education, and changes the standard development process quite drastically \cite{Clarke1996, Abrial2006}.
Furthermore, both boundaries are limited by humans, i.e., the formal model is just as good as the requirements document.
Similarly, the translated code is only as correct as the human implemented translator is.
However, safety-critical systems greatly benefit from FM because of its ability to provide proof of the behavior \cite{Abrial2006}.
Compared to PSM, FM moves the comprehension of the application and its requirements into the design phase when the specification models are created.
PSM may mitigate some of the issues by reducing the FM part to critical systems while non-critical systems are comprehended using PSM.

Symbolic Execution \cite{King1976} avoids structural abstraction level changes by combining testing and formal methods.
It generalizes testing by replacing real variable values, with symbolic values and keeps track of the symbolic program path.
This allows the exploration of multiple program paths simultaneously giving an understanding of which program points are reachable or not.
Typical applications are programs path space comprehension, test input generation, and finding bugs or vulnerabilities of the code.
Challenges are for example path explosions, external environments, and loop invariants \cite{Baldoni2016}.
In comparison to PSM, Symbolic Execution abstracts the concrete values to path conditions while PSM models a distribution of the encountered values.
Symbolic Execution finds inconsistencies in the code by systematically solving for all possible program paths.
PSM, on the other hand, builds distributions of the application components and captures questionable behavior across them.
Probabilistic Program Analysis (PPA) \cite{Filieri2014} is an extension of Symbolic Execution that quantifies program paths and points utilizing probabilities, instead of the binary decision \emph{satisfiable} and \emph{not satisfiable}.
So given a program and a statement of interest, Probabilistic Program Analysis can quantify the likelihood a specific statement is executed.
Regardless of the similar names, PSM and Probabilistic Program Analysis do not share many similarities.
PSM models the content of a statement over multiple executions, while PPA quantifies the likelihood that a statement is executed (PPA).

\section{Research Goals and Evaluation Strategies}

The goal of this work is to devise and define the essential
\begin{enumerate*}
	\item characteristics,
	\item methodologies, and
	\item applications, that characterize probabilistic software modeling.
\end{enumerate*}
These definitions and strategies should guide the implementation of a prototypical PSM system and the introduction of the modeling paradigm itself.
\emph{Characteristics} will be concerned with \emph{when} and \emph{where} PSM is the right choice based on the success of the implemented applications.
\emph{Methodologies} describe how to exploit the benefits and drawbacks of structural and temporal dependencies of the system's random variables (types, properties, executables).
\emph{Applications} provide means to quantify the approach empirically, and its appropriateness in the respective domain.
The combination of the three research goals (characteristics, methodologies, applications) will allow the evaluation of PSM and its capabilities to solve the problem of software comprehension.

The implementation of a prototypical PSM system will drive the evaluation of the research.
PSM tries to estimate the densities of code elements at runtime, hence, is an unsupervised machine learning approach that is typically harder to evaluate as well-known supervised methods.
Benchmarks, such as DaCapo \cite{Blackburn}, will be used to evaluate the prototype on real-world examples measuring the estimated fit of the PMs.
This fit is evaluated by visual inspection (qualitative) of interesting cases (such as strings), and by metrics (quantitative) that measure the error between sampled and original dataset quantifying bias and variance of the models.
On a higher level, the prototype will be evaluated by implementing concrete applications mentioned in Section \ref{sec:applications}.
These will be subject to studies that compare the PSM solution to the current state-of-art.
Both qualitative and quantitative evaluations of the studies mentioned above will provide evidence whether PSM improves on the current state-of-art of supporting program comprehension.
That is, by either improving existing techniques (such as test-case generation) or by additional support (such as inferential queries, simulation).

\section{Conclusion}
We presented \emph{Probabilistic Software Modeling (PSM)}, a complementary software modeling approach.
The goal of PSM is to support software comprehension from a code-centric perspective.
The approach complements MDE in that it provides engineers a way to reason about software and its runtime after a system has been implemented.
Further, this reasoning is done on the same level as the implementation concepts (types, properties, and executables).
An overview of the approach was given on a conceptual level which was then compared with existing other modeling approaches to see commonalities and possible synergies.
At last, we provided an outlook of the research methodology, possible applications, and future studies to evaluate PSM.